\documentclass[12pt]{article}
\usepackage{amssymb, amsmath}
\usepackage{graphicx, longtable}
\usepackage{graphics}

\title{Influence of anisotropic next-nearest-neighbor hopping on
diagonal charge-striped phases}
\author{V. Derzhko\\ Institute of Theoretical Physics, University of Wroc{\l}aw,\\ pl.
Maksa Borna 9, 50--204 Wroc{\l}aw, Poland\\ e-mail:
derzhko@ift.uni.wroc.pl}

\begin{document}

\maketitle
%----------------------------------------------------------------------
\begin{abstract}
We consider the model of strongly-correlated system of electrons
described by an extended Falicov--Kimball Hamiltonian where the
stability of some axial and diagonal striped phases was proved.
Introducing a next-nearest-neighbor hopping, small enough not to
destroy the striped structure, we examine rigorously how the
presence of the next-nearest-neighbor hopping anisotropy reduces the
$\pi/2$-rotation degeneracy of the diagonal-striped phase. The
effect appears to be similar to that in the case of anisotropy of
the nearest-neighbor hopping: the stripes are oriented in the
direction of the weaker next-nearest-neighbor hopping.
\end{abstract}
%----------------------------------------------------------------------

\section{Introduction}
Experimentally, the presence of the striped structures in materials
exhibiting high-temperature superconductivity is well-known
\cite{TBSL1,TSANU1}. Theoretical investigations of the stability and
properties of striped phases are performed mainly with approximate
methods. Usually, the Hubbard-like or $t$--$J$-like models, as the
standard models applied to explain the effects in strongly
correlated systems, are considered for this purpose (see for example
a review \cite{Oles1}). Unfortunately, due to the tiny energy
differences between the energies of compared phases, the methods
that are applied to study the stability are not reliable. That is
why some simpler models (see Refs. \cite{ZH1,HZ1,LFB1,LFB2}) are
considered in the investigations of striped-structures stability and
properties.

In the recent paper \cite{DJ1} we addressed the issue of
striped-order formation in the systems of strongly-correlated
quantum particles described by extended Falicov--Kimball models. We
showed that at half-filling and in the strong-coupling regime some
axial and diagonal striped phases are stable. In comparison to the
standard spinless Falicov--Kimball model (like in
Ref.~\cite{BS1,KL1}) the model in Ref.~\cite{DJ1} was augmented by a
direct Ising-like nearest-neighbor (n.n.) attractive interaction
between the immobile particles, in order to allow for segregated
phases. Changing the intensity of the Ising-like n.n. interaction,
the system is driven from a crystalline (chessboard) phase to a
segregated phase, via quasi-one-dimensional striped phases. This
conclusion was reached for two kinds of hopping particles: fermions
and hard-core bosons.

The obtained results enable us to continue the study of the
properties of striped phases. One of the interesting questions, the
influence of n.n. anisotropy of hopping intensity on axial striped
phases, was investigated by means of the Hartree--Fock method in
\cite{Oles1,RNO1}. In Ref.~\cite{DJ2}, at the regime where stripes
are stable, we have proved rigorously that for both systems, of
hopping fermions and hard-core bosons, an arbitrarily small
anisotropy of n.n. hopping orients the axial striped phases in the
direction of a weaker hopping. We noted also the tendency of the
phase diagrams for different statistics to become similar, even for
a weak anisotropy of n.n. hopping.

\begin{figure}[t]
\centering \includegraphics[width=0.9\textwidth]{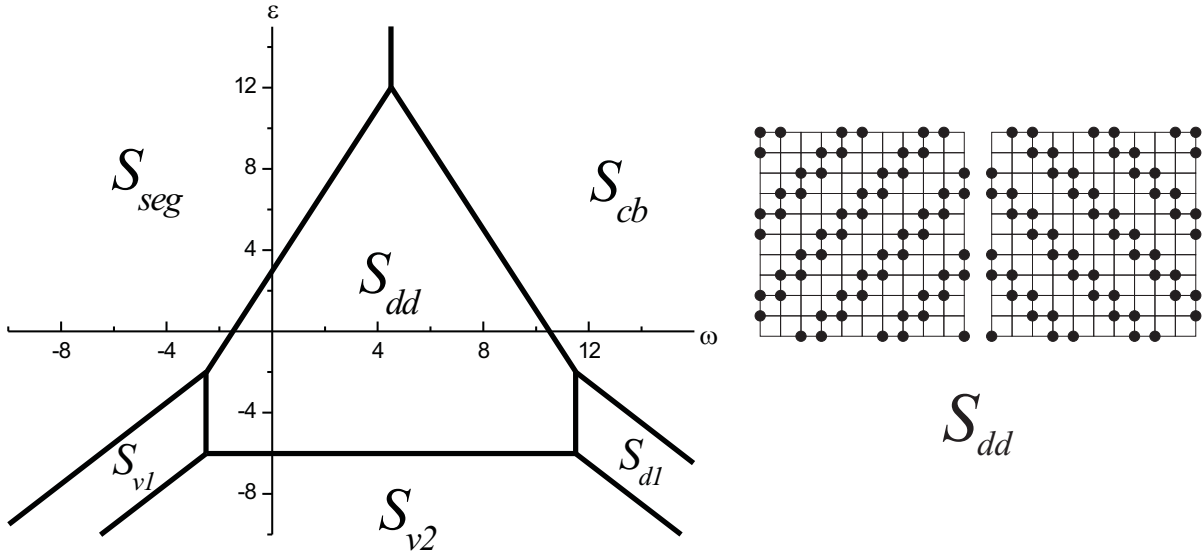}
\caption{The phase diagram obtained in \cite{DJ1}, containing a
diagonal-striped phase $\mathcal{S}_{dd}$, at $\mu=0$, in the plane
$(\omega, \varepsilon)$. In this paper $\omega$ is given by
$W=-2t^{2}+\omega t^{4}$, see (\ref{HV}), and $\varepsilon=0$. The
representative configurations (up to translations) of phase
$\mathcal{S}_{dd}$ are shown on the right. For more details see
\cite{DJ1}.} \label{pdold}
\end{figure}

The analogous, arising naturally, question is how the anisotropy of
the next-nearest-neighbor (n.n.n.) hopping influences the degeneracy
of diagonal-striped phases. To answer this question, in this paper,
we use the same techniques as in \cite{DJ2}. Specifically, we
investigate the influence of n.n.n. hopping on the phase
$\mathcal{S}_{dd}$, whose stability was proved for fermions in
\cite{DJ1} (in Fig.~\ref{pdold} we reproduce the phase diagram, and
show representative configurations of the phase $\mathcal{S}_{dd}$).
According to the state of art, that is possible
only in the {\em{strong-coupling regime}} and at the
{\em{half-filling}}.

In the next section we present the model and discuss some of its
basic properties. In Section 3, we introduce a strong-coupling
expansion of the ground state energy (effective Hamiltonian). In
Section 4, we carry out the analysis of phase diagram due to the
truncated effective Hamiltonian. Finally, we make conclusions in the
Summary.
%----------------------------------------------------------------------

\section{The model and its basic properties}
The model we consider here, is an extension of the spinless
Falicov--Kimball model \cite{KL1}, which in turn is a simplified
version of the one introduced in \cite{FK1}. The system is augmented
by a direct Ising-like interaction between immobile particles.
Hopping particles are allowed to hop not only between n.n. sites (as
is considered usually), but between n.n.n. sites as well. Moreover,
the n.n.n. hopping intensity depends on direction. The Hamiltonian
of the system is of the form:
\begin{align}
H_{0}&=H_{FK}+V, \label{H0} \displaybreak[0] \\
H_{FK}&=- t \sum\limits_{\langle x,y \rangle_{1}} \, \left(
c^{+}_{x}c_{y} + c^{+}_{y}c_{x} \right) -t_{+} \sum\limits_{\langle
x,y \rangle_{2,+}} \, \left( c^{+}_{x}c_{y} + c^{+}_{y}c_{x} \right)
-t_{-} \sum\limits_{\langle x,y \rangle_{2,-}} \, \left(
c^{+}_{x}c_{y} + c^{+}_{y}c_{x} \right) \nonumber
\displaybreak[0] \\
&
+U\sum\limits_{x}\left(c^{+}_{x}c_{x} - \frac{1}{2} \right) s_{x}, \label{HFK}\displaybreak[0] \\
V&=\frac{W}{8} \sum\limits_{\langle x,y \rangle_{1}} s_{x}s_{y}
\label{HV}.
\end{align}
The underlying lattice $\Lambda$ consists of $|\Lambda|$ sites
denoted $x,y,\ldots$, and having the shape of a
$\sqrt{|\Lambda|}\times\sqrt{|\Lambda|}$ torus. The sum
$\sum_{\langle x,y \rangle_{1}}$ stands for summation over all n.n.
pairs, while the sums $\sum_{\langle x,y \rangle_{2,+}}$,
$\sum_{\langle x,y \rangle_{2,-}}$ --- for summation over all n.n.n.
pairs oriented in the direction with slope $+1$ ($+$) or in the
direction with slope $-1$ ($-$), respectively.

The subsystem of quantum hopping particles (here-after called
{\em{the electrons}}) is described in terms of creation and
annihilation operators of an electron at site $x$, $c^{+}_{x}$,
$c_{x}$, respectively, satisfying the canonical anticommutation
relations. The total electron number, $N_{e}$, is the eigenvalue of
$\sum_{x}c^{+}_{x}c_{x}$, and the corresponding electron density is
$\rho_{e}=N_{e}/|\Lambda|$. Although the electrons do not interact
directly with each other, there is an on site interaction with the
localized particles that brings energy $U$ if a site is occupied by
the both kinds of particles.

Since the site-occupation-number operators of classical immobile
particles (here-after called {\em{the ions}}) commute with
Hamiltonian (\ref{H0}), the subsystem of ions can be described by a
set of pseudo-spins $\{ s_{x} \}_{x\in \Lambda}$ (the {\em{ion
configurations}}), with $s_{x}=\pm1$ ($1$ for the occupied site and
$-1$ for the empty site). The total number of ions is
$N_{i}=\sum_{x}(s_{x}+1)/2$ and the ion density is
$\rho_{i}=N_{i}/|\Lambda|$. In our model the ions interact directly
via a n.n. Ising-like interaction $V$, which is isotropic.

The Hamiltonian $H_{FK}$ is the well-known spinless Hamiltonian of
the Falicov--Kimball model. A review of rigorous results and an
extensive list of references concerning this model can be found in
\cite{JL1,GM1,GU1}.

We shall study the ground-state phase diagram of this system using
the grand-canonical formalism, i.e. we consider
\begin{eqnarray}
H\left(\mu_{e}, \mu_{i}\right)=H_{0} - \mu_{e}N_{e} - \mu_{i}N_{i},
\label{HGC}
\end{eqnarray}
where $\mu_{e}$, $\mu_{i}$ are the chemical potentials of electrons
and ions, respectively. Let $E_{S}\left( \mu_{e}, \mu_{i} \right)$
be the ground-state energy of $H \left( \mu_{e}, \mu_{i} \right)$,
for a given configuration $S$ of the ions. Then, the ground-state
energy of $H \left( \mu_{e}, \mu_{i} \right)$, $E_{G}\left( \mu_{e},
\mu_{i} \right)$,  is defined as $E_{G}\left( \mu_{e}, \mu_{i}
\right) = \min \left\{ E_{S} \left( \mu_{e}, \mu_{i} \right): S
\right\}$. The minimum is attained at the set $G$ of the
ground-state configurations of ions.

Applying unitary transformations to $H\left(\mu_{e}, \mu_{i}\right)$
we can restrict the range of the energy parameters. Firstly, using
the hole--particle transformation for ions, $s_{x}\to-s_{x}$, one
finds that the case of attraction ($U>0$) and that of repulsion
($U<0$) are related by this transformation: if
$S=\{s_{x}\}_{x\in\Lambda}$ is a ground-state configuration at
$(t,t_{+},t_{-},\mu_{e},\mu_{i},U)$, then
$-S=\{-s_{x}\}_{x\in\Lambda}$ is the ground-state configuration at
$(t,t_{+},t_{-},\mu_{e},-\mu_{i},-U)$. Consequently, one can fix the
sign of the coupling constant $U$ without any loss of generality. We
choose $U>0$. We  express also all the other parameters of the
Hamiltonian (\ref{HGC}) in the units of $U$, i.e. formally we set
$U=1$, preserving previous notations.

Secondly, applying the hole--particle transformation for electrons,
i.e. $c_{x}\rightarrow \varepsilon_{x}c_{x}^{+}$ (where
$\varepsilon_{x}$ is equal to $+1$ on the even sublattice, and $-1$
on the odd one), and the hole--particle transformation for ions, one
finds that if $S$ is the ground-state configuration at
$(t,t_{+},t_{-},\mu_{e},\mu_{i})$, then $-S$ is the ground-state
configuration at $(t,-t_{+},-t_{-},-\mu_{e},-\mu_{i})$. Thus, it is
enough to consider only one sign of n.n.n. hopping. We shall
consider the case of positive n.n.n. hopping intensities:
$t_{+},t_{-}>0$.

Thirdly, using another hole--particle transformation for electrons,
$c_{x}\rightarrow c_{x}^{+}$, and the hole-particle transformation
for ions, we obtain that if $S$ is the ground-state configuration at
$(t,t_{+},t_{-},\mu_{e},\mu_{i})$, then $-S$ is the ground-state
configuration at $(-t,-t_{+},-t_{-},-\mu_{e},-\mu_{i})$. Applying
consecutively the two joint (with respect to electrons and ions)
hole--particle transformations, we obtain that if $S$ is the
ground-state configuration at $(t,t_{+},t_{-},\mu_{e},\mu_{i})$, it
is also the ground-state configuration at
$(-t,t_{+},t_{-},\mu_{e},\mu_{i})$. So the relative sign of n.n. and
n.n.n. hopping amplitudes does not play any role in the model.

Finally, let us note that Hamiltonian (\ref{HGC}), in contrast to
that where only n.n. hopping is present, is not invariant with
respect to the joint hole--particle transformation for electrons and
ions for any values of $\mu_{e}$ and $\mu_{i}$.
%----------------------------------------------------------------------

\section{The strong-coupling expansion of the ground-state energy}
\begin{figure}
\centering \includegraphics[width=0.9\textwidth]{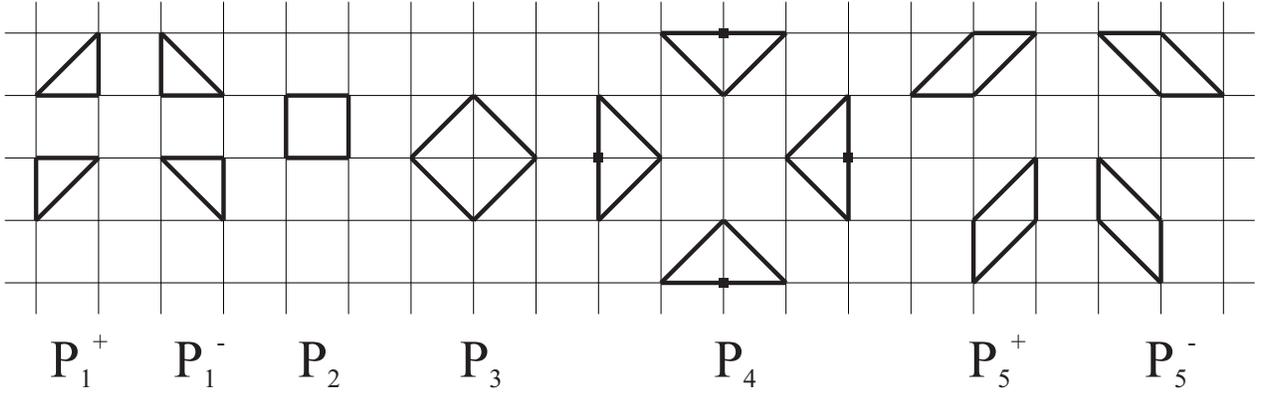}
\caption{The set of plaquettes (paths) over which the sums are taken
in the ground-state energy expansion (\ref{GSEE}).} \label{paths}
\end{figure}

Using the method of unitary-equivalent interactions \cite{DFF1}, in
the strong-coupling regime and at half-filling we can expand the
ground-state energy $E_{S}$ into a power series in
$t^{a}t_{+}^{b}t_{-}^{c}$. The result, with the expansion terms up
to the fourth order (the {\em{fourth-order effective Hamiltonian}}),
i.e. $a+b+c\leqslant4$, reads:
\begin{align}
E_{S}\left( \mu_{e}, \mu_{i} \right) =& E_{S}^{(4)}\left( \mu_{e}, \mu_{i} \right) +
R^{(4)}, \nonumber \\
E_{S}^{(4)}\left( \mu_{e}, \mu_{i} \right) =&
-
\left[
\frac{1}{2}\left( \mu_{i} - \mu_{e} \right)+
\frac{3}{4}t^{2}\left( t_{+}+t_{-} \right)
\right]
\sum\limits_{x}s_{x}
+
\nonumber \displaybreak[0]\\
&+
\left[
\frac{1}{4}t^{2}
-\frac{9}{16}t^{4}
-\frac{1}{16}t^{2}\left( 3t_{+}^{2}+10t_{+}t_{-}+3t_{-}^{2} \right)
+\frac{W}{8}
\right]
\sum\limits_{\langle x,y \rangle_{1}} s_{x}s_{y}
+
\nonumber \displaybreak[0]\\
&+
\left[
\frac{1}{4}t_{+}^{2}
+\frac{3}{16}t^{4}
-\frac{3}{8}t^{2}\left(2t_{+}^{2}+t_{+}t_{-}\right)
-\frac{3}{16}t_{+}^{4}
-\frac{3}{8}t^{2}_{+}t^{2}_{-}
\right]
\sum\limits_{\langle x,y \rangle_{2,+}} s_{x}s_{y}
+
\nonumber \displaybreak[0]\\
&+
\left[
\frac{1}{4}t_{-}^{2}
+\frac{3}{16}t^{4}
-\frac{3}{8}t^{2}\left(2t_{-}^{2}+t_{+}t_{-}\right)
-\frac{3}{16}t_{-}^{4}
-\frac{3}{8}t^{2}_{+}t^{2}_{-}
\right]
\sum\limits_{\langle x,y \rangle_{2,-}} s_{x}s_{y}
+
\nonumber \displaybreak[0]\\
&+
\left[
\frac{1}{8}t^{4}
-\frac{1}{8}t^{2}t_{+}t_{-}
+\frac{3}{16}t^{2}_{+}t^{2}_{-}
\right]
\sum\limits_{\langle x,y \rangle_{3}} s_{x}s_{y}
+
\frac{3}{16}t^{2}t_{+}^{2}
\sum\limits_{\langle x,y \rangle_{4,+}} s_{x}s_{y}
+
\frac{3}{16}t^{2}t_{-}^{2}
\sum\limits_{\langle x,y \rangle_{4,-}} s_{x}s_{y}
+
\nonumber \displaybreak[0]\\
&+
\frac{1}{8}t_{+}^{4}
\sum\limits_{\langle x,y \rangle_{5,+}} s_{x}s_{y}
+
\frac{1}{8}t_{-}^{4}
\sum\limits_{\langle x,y \rangle_{5,-}} s_{x}s_{y}
+
\frac{3}{8}t^{2}t_{+}
\sum\limits_{P^{+}_{1}}s_{P^{+}_{1}}
+
\frac{3}{8}t^{2}t_{-}
\sum\limits_{P^{-}_{1}}s_{P^{-}_{1}}
+
\nonumber\displaybreak[0]\\
&+
\frac{5}{16}
\left[
t^{4}+
2t^{2}t_{+}t_{-}
\right]
\sum\limits_{P_{2}} s_{P_{2}}
+
\frac{5}{16}
t^{2}_{+}t^{2}_{-}
\sum\limits_{P_{3}} s_{P_{3}}
+
\frac{5}{16}
t^{2}t_{+}t_{-}
\sum\limits_{P_{4}} s_{P_{4}}
+
\nonumber\displaybreak[0]\\
&+
\frac{5}{16}
t^{2}t^{2}_{+}
\sum\limits_{P^{+}_{5}} s_{P^{+}_{5}}
+
\frac{5}{16}
t^{2}t^{2}_{-}
\sum\limits_{P^{-}_{5}} s_{P^{-}_{5}}
.
\label{GSEE}
\end{align}
Here, the sign $\bullet$ in the $\langle x,y \rangle_{i,\bullet}$
means a positive or a negative slope of an $i$-th order n.n. pair.
The sets of plaquettes (or paths) $P^{\bullet}_{i}$ are shown in
Fig.~\ref{paths}.
The sign $\bullet$ in the superscript of the path $P^{\bullet}_{i}$
reflects the sign of the slope of the n.n.n. pairs in it. The
remainder $R^{(4)}$ is independent of the chemical potentials and
$W$, and collects all the terms proportional to
$t^{a}t^{b}_{+}t^{c}_{-}$, with $a+b+c=5,6,\ldots$. The above
expansion is absolutely convergent for sufficiently small $t$,
$t_{+}$ and $t_{-}$, uniformly in $\Lambda$. In the special case of
$t_{+}=t_{-}=t^{\prime}$ and $W=0$ it was obtained in
\cite{Wojtkiewicz}.

Let us note that the ground-state energy expansion (\ref{GSEE})
depends only on $(\mu_{i}-\mu_{e})$. Hence, we denote this
difference as the unique chemical potential parameter $\mu$.

In Ref.~\cite{DJ1}, we have obtained the phase diagram of the
isotropic model without n.n.n. hopping, i.e. for $t_{+}=t_{-}=0$.
Here, our aim is to determine the influence of the n.n.n.-hopping
anisotropy on the diagonal-striped phase $\mathcal{S}_{dd}$ (see
Fig.~\ref{pdold}). For this job, the value of the n.n.n.-hopping
intensities, $t_{+}$, $t_{-}$, cannot be too large, in order to
preserve the phase diagram up to 4th order. On the other hand, the
n.n.n.-hopping intensities cannot be too small, in order to appear
in the fourth-order effective Hamiltonian. In an attempt to satisfy
the both requirements, we choose the smallest n.n.n.-hopping
intensities $t_{+}$, $t_{-}$, i.e. such that they do not appear in
the expansion terms of order smaller than four: $t_{+}=a_{+}t^{2}$
and $t_{-}=a_{-}t^{2}$. In this case, the effective Hamiltonian
assumes the form:
\begin{align}
E_{S}^{(4)}\left( \mu \right) =&
-
\left[
\frac{1}{2}\mu+
\frac{3}{4}t^{4}\left( a_{+}+a_{-} \right)
\right]
\sum\limits_{x}s_{x}
+
\left[
\frac{1}{4}t^{2}
-\frac{9}{16}t^{4}
+\frac{W}{8}
\right]
\sum\limits_{\langle x,y \rangle_{1}} s_{x}s_{y}
+
\nonumber \displaybreak[0]\\
&+
\left[
\frac{1}{4}t^{4}a_{+}^{2}
+\frac{3}{16}t^{4}
\right]
\sum\limits_{\langle x,y \rangle_{2,+}} s_{x}s_{y}
+
\left[
\frac{1}{4}t^{4}a_{-}^{2}
+\frac{3}{16}t^{4}
\right]
\sum\limits_{\langle x,y \rangle_{2,-}} s_{x}s_{y}
+
\nonumber \displaybreak[0]\\
&+
\frac{1}{8}t^{4}
\sum\limits_{\langle x,y \rangle_{3}} s_{x}s_{y}
+
\frac{3}{8}t^{4}a_{+}
\sum\limits_{P^{+}_{1}}s_{P^{+}_{1}}
+
\frac{3}{8}t^{4}a_{-}
\sum\limits_{P^{-}_{1}}s_{P^{-}_{1}}
+
\frac{5}{16}
t^{4}
\sum\limits_{P_{2}} s_{P_{2}},
\label{Haa}
\end{align}
i.e. the second requirement is satisfied. To answer the question put
in the introduction, concerning the influence of anisotropy of
n.n.n. hopping on the degeneracy of the phase $\mathcal{S}_{dd}$,
there is no need to consider the whole phase diagram. For
$t_{+}=t_{-}=0$, we fix a point, well inside the domain of the
diagonal-striped phase $S_{dd}$, say $\mu=0$ and
$W=-2t^{2}+9/2t^{4}$, i.e. $\omega=9/2$ (see Fig.~\ref{pdold}).
Then, with the fixed point in $(\mu,W)$-plain, we introduce a n.n.n.
hopping which does not change the ground-state configurations.
Calculations show that $\mathcal{S}_{dd}$ has the minimal energy for
$a=|a_{+}|=|a_{-}|$, where $-1/4\leqslant a\leqslant1/4$ (we suppose
that the difference between $a_{+}$ and $a_{-}$ is not large, so
they are of the same sign). Therefore, with our choice of
n.n.n.-hopping intensities, the first of the above two requirements
can also be satisfied. Eventually, we fix the values of
n.n.n.-hopping intensities: $a_{+}=1/8$, $a_{-}=\gamma a_{+}$, with
$\gamma$ varying about $1$ (say, $0\leqslant\gamma\leqslant2$).

Since all the energy parameters, except the parameter $\gamma$ of
n.n.n.-hopping anisotropy, have been fixed, the effective
Hamiltonian (\ref{Haa}) depends only on $\gamma$. In the following
section, we examine how n.n.n.-hopping anisotropy, $\gamma \neq 1$,
influences the degeneracy of the diagonal-striped phase
$\mathcal{S}_{dd}$.
%----------------------------------------------------------------------

\section{Diagonal-striped phase versus n.n.n.-hopping anisotropy}
We use the $m$-potential method \cite{Slawny1} for constructing the
phase diagram of effective Hamiltonian $E_{S}^{(4)}(\gamma)$. For
technical reasons, it is convenient to deal with such energies of
configurations that are affine functions of the parameters of the
effective Hamiltonian. However, the effective Hamiltonian
(\ref{Haa}) contains the terms proportional to $\gamma$ and
$\gamma^{2}$. To get rid of nonlinearities, we replace $\gamma$ and
$\gamma^{2}$ by two independent parameters $d_{1}$ and $d_{2}$,
respectively, with $d_1,d_2$ varying in the rectangle $0\leq d_1
\leq 2$ and $0\leq d_2 \leq 4$, in which the Hamiltonian is affine.
After constructing the phase diagram in $(d_{1},d_{2})$-plane, we
restrict it to the $d_{2}=d^{2}_{1}$ curve.

To compare the energies of configurations, we rewrite
$E_{S}^{(4)}(d_{1},d_{2})$ as the sum,
\begin{eqnarray}
E_{S}^{(4)}\left( d_{1},d_{2}
\right)=\frac{t^{4}}{2}\sum\limits_{T}H^{(4)}_{T}, \label{Epot}
\end{eqnarray}
over $(3\times 3)$-site blocks (called $T$-plaquettes). The
potential $H^{(4)}_{T}$ is of the form:
\begin{eqnarray}
H^{(4)}_{T}&=&-\frac{3}{16}\left(d_{1}+1\right)s_{5} +
\frac{49}{512}\sideset{}{^{\prime\prime}}\sum\limits_{\langle x,y
\rangle_{2,+}}s_{x}s_{y} + \frac{1}{32}\left( \frac{1}{16}d_{2}+3
\right)\sideset{}{^{\prime\prime}}\sum\limits_{\langle x,y
\rangle_{2,-}}s_{x}s_{y}+\frac{1}{12}\sideset{}{^{\prime\prime}}\sum\limits_{\langle
x,y \rangle_{3}}s_{x}s_{y}+\nonumber\\
&&+\frac{3}{128}\sideset{}{^{\prime\prime}}\sum\limits_{P^{+}_{1}}s_{P^{+}_{1}}
+\frac{3}{128}d_{1}\sideset{}{^{\prime\prime}}\sum\limits_{P^{-}_{1}}s_{P^{-}_{1}}
+\frac{5}{32}\sideset{}{^{\prime\prime}}\sum\limits_{P_{2}}s_{P_{2}},
\label{HT4}
\end{eqnarray}
where $s_{5}$ is the central site of a $T$-plaquette.

Unfortunately, the potential (\ref{HT4}) is not an $m$-potential in
the rectangle of considered values of $d_1$ and $d_2$. Therefore,
following \cite{GJL1,GMMU1,Kennedy1} we introduce so-called
{\em{zero-potentials}}. The zero potentials, satisfying
\begin{eqnarray}
\sum\limits_{T}K_{T}^{(4)}=0, \label{zerpotcond}
\end{eqnarray}
can be chosen in the form:
\begin{eqnarray}
K_{T}=\sum\limits_{i=1}^{9}\alpha_{i}k_{T}^{(i)},
\end{eqnarray}
where coefficients $\alpha_{i}$ have to be determined in the process
of constructing a phase diagram, and the potentials $k_{T}^{(i)}$,
invariant with respect to the spatial symmetries of $H_{0}$, and
fulfilling (\ref{zerpotcond}), read:
\begin{align*}
k_{{\rm{T}}}^{(1)} = & s_{1} +s_{9} -2s_{5}, \displaybreak[0] \\
k_{{\rm{T}}}^{(2)} = & s_{2} +s_{8} -2s_{5}, \displaybreak[0] \\
k_{{\rm{T}}}^{(3)} = & s_{3} +s_{7} -2s_{5}, \displaybreak[0] \\
k_{{\rm{T}}}^{(4)} = & s_{4} +s_{6} -2s_{5}, \displaybreak[0] \\
k_{{\rm{T}}}^{(5)} = & s_{1}s_{2} +s_{8}s_{9} -s_{4}s_{5} -s_{5}s_{6},\displaybreak[0] \\
k_{{\rm{T}}}^{(6)} = & s_{2}s_{3} +s_{7}s_{8} -s_{4}s_{5} -s_{5}s_{6},\displaybreak[0] \\
k_{{\rm{T}}}^{(7)} = & s_{1}s_{4} +s_{6}s_{9} -s_{2}s_{5} -s_{5}s_{8},\displaybreak[0] \\
k_{{\rm{T}}}^{(8)} = & s_{3}s_{6} +s_{4}s_{7} -s_{2}s_{5} -s_{5}s_{8},\displaybreak[0] \\
k_{{\rm{T}}}^{(9)} = & s_{2}s_{4} +s_{6}s_{8} -s_{3}s_{5}
-s_{5}s_{7}.
\end{align*}
Here, we label $1,\ldots,9$, the sites of a $T$-plaquette, from left
to right, starting in the bottom left corner and ending in the upper
right one. In order to obtain the phase diagram, we have to compare
the energies of all the possible $T$-plaquette configurations. The
zero-potential coefficients $\alpha$ needed for this are given in
Tab.~\ref{tab1} in the Appendix. We provide their values only at
certain generating points, since we can assume that the coefficients
$\alpha$ are affine functions of parameters $(d_{1},d_{2})$. For
more details on using the $m$-potential method see \cite{DJ2} and
references therein.

\begin{figure}[t]
\centering \includegraphics[width=0.4\textwidth]{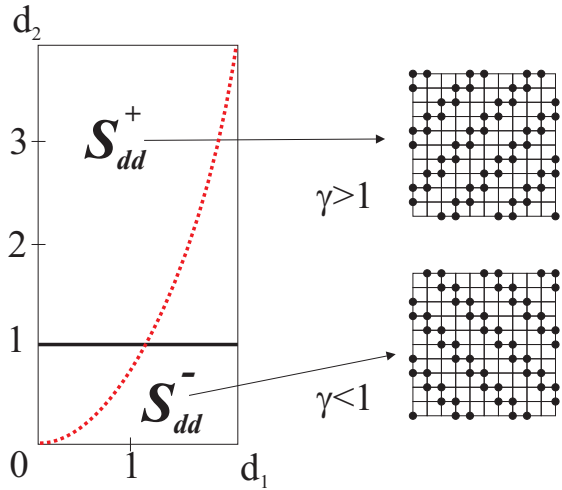}
\caption{The phase diagram of $E^{(4)}(d_{1},d_{2})$, with
$0\leqslant d_{1}\leqslant2$ and $0\leqslant d_{2}\leqslant4$. The
rectangle breaks down into two regions: above $d_{2}=1$ the
configurations $\mathcal{S}^{+}_{dd}$ are the ground-state
configurations, while $\mathcal{S}^{+}_{dd}$ are replaced by
$\mathcal{S}^{-}_{dd}$ below $d_{2}=1$. The dashed line represents
the condition $d_{2}=d_{1}^{2}$. The representative configurations
(up to translations) of $\mathcal{S}^{+}_{dd}$,
$\mathcal{S}^{-}_{dd}$ are shown on the right.} \label{pdd1d2}
\end{figure}

The phase diagram of $E^{(4)}$ is shown in Fig.~\ref{pdd1d2}.
The rectangle of considered points $(d_{1},d_{2})$ breaks down into
two domains. In the lower one, where $\gamma<1$ and $t_{-}<t_{+}$,
it is the phase $\mathcal{S}^{-}_{dd}$, with stripes being parallel
to the direction of $t_{-}$-hopping, that is stable. The analogous
situation is in the upper domain, where $\gamma>1$ and
$t_{-}>t_{+}$: the stable phase, $\mathcal{S}^{+}_{dd}$, consists of
stripes oriented along $t_{+}$-hopping. At $\gamma =1$, we have the
isotropic phase $\mathcal{S}_{dd}$ whose configurations consist of
$\mathcal{S}^{+}_{dd}$ and $\mathcal{S}^{-}_{dd}$.

So we see that, at least for the truncated effective Hamiltonian,
switching on of a n.n.n.-hopping anisotropy reduces the rotational
degeneracy of diagonal-striped phases: they become oriented in the
direction of the weaker hopping.

This result is similar to that described in \cite{DJ2}, where the
influence of n.n.-hopping anisotropy on axial-striped phases was
investigated. In that case, not only for a truncated effective
Hamiltonian but also for the corresponding quantum one, it was
proved that for any nonzero value of n.n.-hopping anisotropy the
rotational degeneracy of axial-striped phases is reduced by making
them oriented along the direction of the weaker n.n. hopping. Now in
turn, the natural question is whether the conclusions we arrived at,
concerning n.n.n.-hopping anisotropy, hold true for the quantum
model, described by Hamiltonian (\ref{HGC}). Applying the arguments
presented in \cite{GMMU1,Kennedy1}, we can demonstrate (see for
instance Ref. \cite{DJ3}), that the stable phases of the obtained
above phase diagram remain stable for the model (\ref{HGC}), but in
some smaller domains. That is, if the remainder $R^{(4)}$ is taken
into account, then there exist such a small $t_{0}$, that for
$t<t_{0}$ the phase diagram looks the same for the quantum model,
excepting of some narrow regions (of width $O(t)$ in the scale of
the fourth-order phase diagram shown in Fig.~\ref{pdd1d2}), located
along the phase-boundary lines. In our case that means that the
breaking of the rotational symmetry occurs for $\gamma=1+O(t)$, when
the n.n.n.-hoping intensities are $O(t^{2})$. Unfortunately, we
cannot claim that any non-zero n.n.n.-hopping anisotropy reduces the
rotational degeneracy of the quantum model, as it was the case for
n.n. hopping (see \cite{DJ2}). Here it seems, at least for small
n.n.n.-hopping intensities, that there is certain critical value of
$|\gamma - 1|$, above which the degeneracy of phase
$\mathcal{S}_{dd}$ is reduced.

%----------------------------------------------------------------------

\section{Summary}
We have considered the model of correlated spinless fermions,
described by an extended Falicov-Kimball Hamiltonian. Quantum
particles are allowed to hop between nearest-neighbor and
next-nearest-neighbor sites. We have shown that a weak anisotropy of
the next-nearest-neighbor hopping reduces the degeneracy of a
diagonal-striped phase, it orients the stripes in the direction of
the weaker next-nearest-neighbor hopping.
%----------------------------------------------------------------------

\section*{Acknowledgments}
The author appreciates Prof. Janusz J{\c{e}}drzejewski's valuable
and fruitful discussions as well as critical review of the
manuscript and thoughtful suggestions.

The author is grateful to University of Wroc{\l}aw for Scientific
Research Grant 2479/W/IFT, and to the Institute of Theoretical
Physics for financial support. The Max Born Scholarship is
gratefully acknowledged.
%----------------------------------------------------------------------

%----------------------------------------------------------------------

\section*{Appendix}
Here we present the zero-potential coefficients $\alpha_{i}$ in the
generating points of phase diagram shown in Fig.~\ref{pdd1d2}.

\renewcommand\baselinestretch{1,5}\small\normalsize

\begin{table}[h]
\begin{center}
\small \caption{Zero-potentials coefficients for the phase diagram
shown in Fig.~\ref{pdd1d2}.} \label{tab1}
\begin{tabular}{|l|c|c|c|c|c|c|}
\hline Points $(d_{1},d_{2})$ & $(0,0)$ & $(2,0)$ & $(0,1)$ &
$(2,1)$ & $(0,4)$ & $(2,4)$
\\
\hline $\alpha_{1}$ & $-\frac{83}{1536}$ & $-\frac{259}{3072}$ &
$-\frac{79}{1536}$ & $-\frac{253}{3072}$ & $-\frac{67}{1536}$ &
$-\frac{217}{3072}$
\\
\hline $\alpha_{2}$ & $-\frac{25}{1536}$ & $-\frac{245}{3072}$ &
$-\frac{29}{1536}$ & $-\frac{251}{3072}$ & $-\frac{41}{1536}$ &
$-\frac{287}{3072}$
\\
\hline $\alpha_{3}$ & $-\frac{47}{1536}$ & $-\frac{331}{3072}$ &
$-\frac{43}{1536}$ & $-\frac{325}{3072}$ & $-\frac{31}{1536}$ &
$-\frac{289}{3072}$
\\
\hline $\alpha_{4}$ & $-\frac{25}{1536}$ & $-\frac{245}{3072}$ &
$-\frac{29}{1536}$ & $-\frac{251}{3072}$ & $-\frac{41}{1536}$ &
$-\frac{287}{3072}$
\\
\hline $\alpha_{5}$ & $\frac{53}{3072}$ & $\frac{137}{8192}$ &
$\frac{19}{1024}$ & $\frac{139}{8192}$ & $\frac{3}{128}$ &
$\frac{91}{8192}$
\\
\hline $\alpha_{6}$ & $-\frac{53}{3072}$ & $-\frac{335}{24576}$ &
$-\frac{19}{1024}$ & $-\frac{365}{24576}$ & $-\frac{3}{128}$ &
$-\frac{91}{8192}$
\\
\hline $\alpha_{7}$ & $\frac{53}{3072}$ & $\frac{401}{8192}$ &
$\frac{19}{1024}$ & $\frac{403}{8192}$ & $\frac{3}{128}$ &
$\frac{1117}{24576}$
\\
\hline $\alpha_{8}$ & $-\frac{53}{3072}$ & $-\frac{571}{24576}$ &
$-\frac{19}{1024}$ & $-\frac{589}{24576}$ & $-\frac{3}{128}$ &
$-\frac{139}{6144}$
\\
\hline $\alpha_{9}$ & $-\frac{19}{3072}$ & $\frac{101}{2048}$ &
$-\frac{19}{3072}$ & $\frac{13}{256}$ & $-\frac{19}{3072}$ &
$\frac{1451}{24576}$
\\
\hline
\end{tabular}
\end{center}
\end{table}
\renewcommand\baselinestretch{1}\small\normalsize
%----------------------------------------------------------------------

\end{document}